\documentclass[aps,reprint,superscriptaddress,notitlepage,nofootinbib,floatfix]{revtex4-2}

\usepackage{amsmath,amssymb}
\usepackage{bm}
\usepackage{graphicx}
\usepackage{xcolor}
\usepackage[hidelinks]{hyperref}

\setcitestyle{super,comma}
\makeatletter
\renewcommand{\section}{\@startsection{section}{1}{\z@}
  {-3ex plus -1ex minus -.2ex}{1.2ex plus .2ex}
  {\normalfont\large\sffamily\bfseries\raggedright}}
\renewcommand{\subsection}{\@startsection{subsection}{2}{\z@}
  {-2.5ex plus -1ex minus -.2ex}{.8ex plus .2ex}
  {\normalfont\normalsize\sffamily\bfseries\raggedright}}
\renewcommand{\subsubsection}{\@startsection{subsubsection}{3}{\z@}
  {-2ex plus -.8ex minus -.2ex}{.6ex plus .2ex}
  {\normalfont\normalsize\itshape\raggedright}}
\renewcommand{\@hangfrom@section}[3]{\@hangfrom{#1#2}#3}
\renewcommand{\@hangfroms@section}[2]{#1#2}

\renewcommand{\fnum@figure}{\textbf{Fig.~\thefigure}}
\renewcommand{\@caption@fignum@sep}{\textbf{~\textbar~}}
\makeatother

\begin{document}

\title{OpenFerro: a unified framework for mesoscale ferroic dynamics and enhanced sampling}

\author{Pinchen Xie}
\email[Correspondence: ]{pinchenxie@lbl.gov}
\affiliation{Applied Mathematics and Computational Research Division, Lawrence Berkeley National Laboratory, Berkeley, California 94720, USA}

\begin{abstract}
Modeling ferroic materials requires connecting microscopic fluctuations to collective dynamics over mesoscales. We introduce OpenFerro, a unified lattice-field dynamics framework that combines flexible Hamiltonians, automatic differentiation, parallel execution, and enhanced sampling. The same framework can simulate ferroelectric domain motion, magnetic disordering, and coupled spin-lattice thermodynamics. On four graphics processors, a bulk BaTiO$_3$ benchmark with long-range dipolar interaction reaches $840^3$ unit cells, representing 3 billion atoms through the retained lattice variables. We simulate polarization textures in a PbTiO$_3$/SrTiO$_3$ superlattice spanning $0.8~\mu\mathrm{m}$ laterally and a BaTiO$_3$ nanoparticle 200~nm in diameter, retaining unit-cell resolution. We also demonstrate enhanced-sampling simulations that reconstruct the polarization free-energy surface of bulk BaTiO$_3$ without a prescribed polynomial form. With moderate computational resources, OpenFerro achieves atomistic-resolution simulations at mesoscopic scales traditionally studied with continuum models.
\end{abstract}

\maketitle

\section{Introduction}

Collective dynamics in ferroic materials, such as domain motions and phase transitions, emerge from interactions among local degrees of freedom. Polarization, magnetic moments, strain, and molecular orientation illustrate the range of variables that can carry these collective behaviors. A domain pattern may span hundreds of nanometers, while its evolution depends on processes confined to a few unit cells at a sharp interface. Modeling such systems requires a description that connects microscopic fluctuations to the evolution of the mesoscopic structures.

Electronic-structure calculations establish microscopic energetics, and atomistic dynamics connect these energies to finite-temperature behavior\cite{hohenberg1964,car1985}. Machine-learned interatomic potentials have substantially extended accessible system sizes\cite{behler2007,zhang2018,batatia2022,song2024}, but evolving every atomic coordinate remains costly when millions of unit cells and slow collective events must be sampled. For mesoscopic domain evolution, continuum phase-field models have long been a principal computational tool\cite{chen2002,karpov2017,das2019}. They describe order through smooth fields and constitutive free energies, enabling extensive studies of nanostructures under varied boundary conditions. Their coarse-graining places local lattice physics into effective coefficients, however, which can limit the description of processes controlled by discrete disorder, sharp interfaces, or local rearrangements.

Lattice-field models provide an intermediate resolution by retaining a few physically motivated variables in each crystal cell and integrating out the remaining coordinates. Such models can retain lattice-scale correlations, long-range interactions, and couplings between different orders. The first-principles effective Hamiltonian of BaTiO$_3$ established this approach for ferroelectric phase transitions\cite{zhong1995}, with subsequent extensions to alloys and coupled structural and magnetic degrees of freedom\cite{bellaiche2000,wang2012multiferroics}. More generally, although rarely demonstrated, temperature-dependent effective Hamiltonians may be defined by integrating the atomistic equilibrium distribution at fixed collective variables\cite{xie2026fes}. This provides a systematic basis for coarse-grained lattice models that can bridge the gap between first-principles and mesoscale simulations.

The reach of lattice-field modeling depends strongly on its implementation. Dedicated ferroelectric and atomistic-spin packages have enabled extensive applications\cite{nishimatsu2008,skubic2008,evans2014}, while numerical optimization and advances in model construction have further expanded their capabilities\cite{prokhorenko2018,ma2025}. Here, a central challenge is that effective Hamiltonians are not unique: their variables and interactions depend on the material and the physical processes of interest. Software built around a particular Hamiltonian often provides limited support for constructing new models, so adding couplings can require changes to force routines and parallel execution. A general framework, therefore, needs to make Hamiltonians straightforward to assemble and extend while providing reusable support for dynamics and enhanced sampling.

Here we introduce OpenFerro, a JAX-based framework for classical lattice-field dynamics. In OpenFerro, a Hamiltonian model is assembled from fields and modular scalar energy functions. Automatic differentiation generates forces on every participating field, so each physical coupling needs to be specified only once. The same Hamiltonian supports structural relaxation, thermal dynamics, and collective-variable biasing, while JAX distributes the lattice fields and force calculations across GPUs\cite{jax2018}. This separation of material physics from simulation and parallelization code allows structural, polar, magnetic, elastic, and orientational variables to share one implementation. 

Our results establish OpenFerro as a general tool for thermodynamics across ferroic orders and their couplings, while demonstrating how scalable GPU execution and enhanced sampling address the length- and time-scale limitations of lattice dynamics. We demonstrate these capabilities using ferroelectric BaTiO$_3$, ferromagnetic bcc Fe, spin-lattice-coupled SrMnO$_3$, and PbTiO$_3$/SrTiO$_3$ superlattices, covering thermodynamic evolution, domain formation, and enhanced free-energy sampling. On a single computing node with four GPUs, we benchmark bulk BaTiO$_3$ dynamics with up to $840^3$ unit cells, representing 3 billion atoms through the retained lattice variables. We also demonstrate simulations of a PbTiO$_3$/SrTiO$_3$ superlattice spanning $0.8~\mu\mathrm{m}$ laterally and a BaTiO$_3$ nanoparticle 200~nm in diameter, retaining a local dipole in every cell. Finally, we use metadynamics to accurately reconstruct the three-dimensional polarization free-energy surface of BaTiO$_3$ across its four phases without assuming a Landau polynomial. This reach makes material-specific lattice dynamics a practical option for nanostructures traditionally studied with continuum phase-field models, allowing whole domain patterns and the local structures of walls, interfaces, and vortex cores to be examined together on a first-principles basis. 

\section{Results}

\subsection{General formulation of classical lattice-field dynamics}

Order parameters and their symmetry-allowed couplings provide a common language for ferroic matter\cite{aizu1970,wadhawan2000}. This perspective also underlies dynamical models that evolve different degrees of freedom together, as demonstrated by coupled spin-lattice simulations of multiferroics\cite{wang2012multiferroics}. OpenFerro builds on this foundation by representing material-specific collective variables as interacting fields on a crystal lattice.

The fields are chosen according to the material and the processes of interest. For example, in a displacive ferroelectric, local soft modes describe the amplitude and direction of cell polarization, while acoustic displacements and homogeneous strain describe local deformation and macroscopic shape. Magnetic moments can be represented by fixed-length vectors, and molecular order by orientational fields. Each lattice site thus carries a set of physically motivated local states tied to the crystal topology, and the interactions among these states are specified by a common Hamiltonian.

OpenFerro represents these states on a three-dimensional Bravais lattice with sites $i=1,\ldots,N$. A site can carry several Euclidean fields $\bm{u}^{(a)}_i\in\mathbb{R}^{d_a}$ and also fixed-magnitude 3D vector fields $\bm{m}_i$. Global variables such as homogeneous strain $\bm{\eta}\in\mathbb{R}^{6}$ coexist with the local fields. The resolution is set by the coarse-graining, usually one primitive or conventional cell. A material Hamiltonian is written as a sum of physically named contributions,
\begin{equation}
  H(\{\bm{u}^{(a)}\},\bm{m},\bm{\eta})
  = \sum_{\alpha} H_{\alpha}(\phi_{\alpha,1},\ldots,
  \phi_{\alpha,n_{\alpha}};\bm{\theta}_{\alpha}),
  \label{eq:hamiltonian}
\end{equation}
where $\phi_{\alpha,j}$ denotes the $j$th of the $n_\alpha$ fields entering $H_\alpha$, and $\bm{\theta}_\alpha$ contains its parameters. Each $\phi_{\alpha,j}$ represents either a complete lattice field, such as $\bm{u}^{(a)}$ or $\bm{m}$, or a global variable, such as $\bm{\eta}$.  

For every Euclidean field, the generalized force follows from the same definition,
\begin{equation}
  \bm{F}^{(a)}_i=-\frac{\partial H}{\partial \bm{u}^{(a)}_i}.
  \label{eq:force}
\end{equation}

Time evolution of real-valued inertial fields can be described by Langevin-equation-based molecular dynamics\cite{bussi2007langevin}. In the canonical ensemble,
\begin{equation}
  M_a\ddot{\bm{u}}^{(a)}_i
  =\bm{F}^{(a)}_i-\gamma_a M_a\dot{\bm{u}}^{(a)}_i
  +\sqrt{2\gamma_a M_a k_{\mathrm B}T}\,\bm{\xi}^{(a)}_i(t),
  \label{eq:langevin}
\end{equation}
where the effective mass ($M_a$) and friction ($\gamma_a$) are part of the coarse-grained dynamical model. A fixed-magnitude magnetic moment evolves under stochastic Landau-Lifshitz dynamics\cite{eriksson2017,mentink2010},
\begin{align}
  \dot{\bm{m}}_i={}&-\gamma_r\bm{m}_i\times(\bm{B}_i+\bm{b}_i) \nonumber\\
  &-\gamma_r\alpha\bm{m}_i\times
  [\bm{m}_i\times(\bm{B}_i+\bm{b}_i)],
  \label{eq:llg}
\end{align}
with 
\begin{equation}\label{eq:mag-force}
  \bm{B}_i=-\frac{\partial H}{\partial\bm{m}_i}
\end{equation}
and thermal noise ($\bm{b}_i$) fixed by the fluctuation-dissipation relation\cite{brown1963}. $\gamma$ is the gyromagnetic ratio, $\gamma_r=\gamma/(1+\alpha^2)$ is its renormalized value, and $\alpha$ is the Gilbert damping constant.

These equations for coarse-grained lattice fields do not imply that time here is automatically an atomistic clock as that in the standard Born-Oppenheimer molecular dynamics. If one wants to compare the real-time correlation functions predicted by these equations with the experimental data, one needs to properly calibrate the effective masses, friction, and damping.

Additionally, for modeling a system with variable volume and for performing constant-pressure simulations, one needs to add $pV(\bm{\eta})$ to the force-generating potential, with $V(\bm{\eta})=V_0\det[\bm{I}+\bm{\epsilon}(\bm{\eta})]$, where $\bm{\epsilon}(\bm{\eta})$ is the symmetric $3\times3$ strain tensor corresponding to the six-component engineering-Voigt vector $\bm{\eta}$.

\subsection{Implementation of lattice-field dynamics through modular automatic differentiation}

\begin{figure}[!htbp]
  \centering
  \includegraphics[width=\columnwidth]{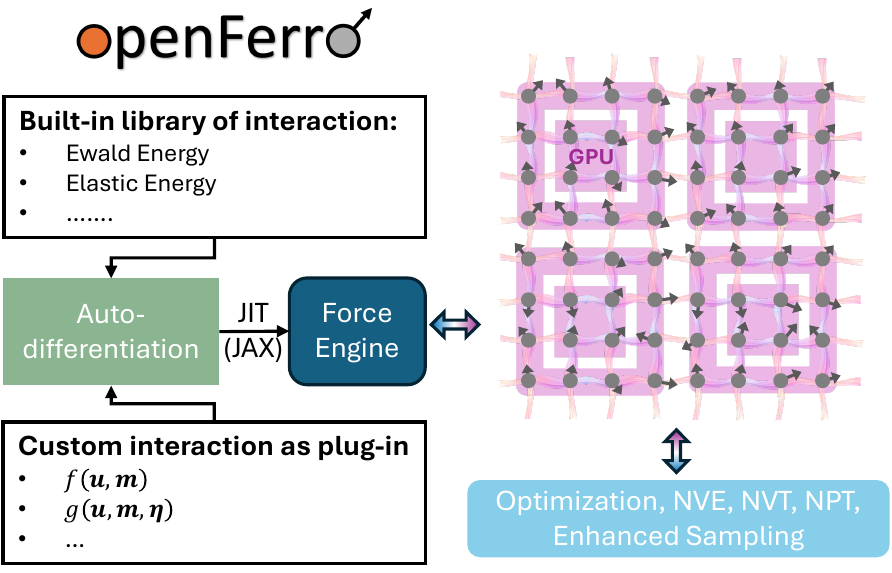}
  \caption{\textbf{Software architecture of OpenFerro.} Built-in interactions and custom energy functions (left) enter a common automatic-differentiation and just-in-time compilation pipeline to generate the force engine. The custom functions can couple local modes $\bm{u}$, spins $\bm{m}$, and strain $\bm{\eta}$. Lattice fields are distributed across GPUs (right), while simulation protocols (bottom) select structural optimization, NVE, NVT, NPT, or enhanced sampling. Material-specific energy functions remain independent of the simulation core.}
  \label{fig:software_architecture}
\end{figure}

Figure~\ref{fig:software_architecture} summarizes how OpenFerro turns a material Hamiltonian into a simulation. Built-in and custom energies pass through a common automatic-differentiation and compilation pipeline to generate forces on lattice fields. The resulting model supports structural relaxation, lattice dynamics, and enhanced sampling, with parallel execution handled by distributing the fields.

% Obtaining the forces in Eq.~\eqref{eq:force} and Eq.~\eqref{eq:mag-force} through automatic differentiation preserves variational consistency across coupled fields. For example, a spin-strain interaction generates both magnetic effective fields and forces on strain from the same energy term. This consistency is retained when new interactions are added to the model.

\subsubsection{Hamiltonian terms as independent modules}

Effective Hamiltonians of a material typically require various anharmonic interactions between orders and interface-specific interactions. OpenFerro treats each interaction term as an independent module that users can add without modifying the package source.

OpenFerro treats built-in interaction libraries and user-defined interaction libraries with the same interface. Users define a Python function mapping lattice-field values and material parameters to a scalar energy through differentiable JAX array operations\cite{jax2018}. For example, a single-field self-interaction term is a Python function in the form of \texttt{energy(field, parameters)}. Registering the function with its participating fields adds it to the Hamiltonian. OpenFerro then differentiates the energy with respect to each field and compiles the corresponding force calculations.

Energy functions can be shared as material libraries together with parameters, units, and reference observables. Material-specific interactions remain independent of the simulation core, allowing models to be extended and reused without separate force derivations or separately maintained versions of OpenFerro.

Specialized operations such as dipolar Ewald summation remain built-in core modules because their implementation and validation require substantial numerical expertise\cite{wang2019ewald}. These operations can be maintained independently and reused across material libraries.

\subsubsection{Automatic parallelism across GPUs and nodes}

OpenFerro distributes the lattice fields across a mesh of GPUs (Fig.~\ref{fig:software_architecture}, right). JAX partitions the energy and force calculations and handles communication. The mesh can span multiple nodes once JAX's distributed runtime is initialized. Custom energy functions inherit this execution model without additional GPU or communication implementations. 

For models with long-range dipolar interactions, the global Fourier transforms in Ewald summation currently impose substantial communication costs. The benchmarks in the following sections therefore focus on single-node multi-GPU performance; cross-node Ewald execution remains experimental. %Improvements to this core module can benefit existing material libraries without changes to their energy functions.

\subsubsection{Unbiased simulations and enhanced sampling}

OpenFerro supports structural optimization, microcanonical (NVE), constant-temperature (NVT), or constant-temperature and pressure (NPT) dynamics. The simulation driver coordinates force updates, field evolution, and reporting. The same material model can thus be used for relaxation, cooling, heating, or field-driven switching protocols.

For flexible Euclidean lattice fields, NVE dynamics use leapfrog integration, while NVT and NPT dynamics use Langevin splitting with velocities stored at half steps\cite{zhang2019thermostat}. Fixed-magnitude fields, such as magnetic spins, evolve with conservative, damped, or stochastic variants of the semi-implicit scheme\cite{mentink2010}. Structural optimization uses force-based gradient descent.

Enhanced sampling (see later sections for details) uses the same machinery. A collective variable, such as a polarization component or a measure of magnetic order, is defined as a differentiable scalar function of the lattice fields. A bias on this variable produces forces on all participating fields through the chain rule. 

OpenFerro supports umbrella sampling and metadynamics\cite{torrie1977,laio2002} in NVT and NPT simulations. Both methods enable the exploration of metastable states and free-energy barriers relevant to domain nucleation and competing ferroic phases.

\subsubsection{Relation to existing simulation packages}

Several established packages have scopes that overlap with OpenFerro. FERAM implements effective-Hamiltonian dynamics for ferroelectrics\cite{nishimatsu2008}, while MULTIBINIT supports first-principles-derived lattice and spin models\cite{verstraete2025}. UppASD\cite{skubic2008} and VAMPIRE\cite{evans2014} provide atomistic spin simulations. Coupled spin-lattice dynamics is also available in MULTIBINIT\cite{multibinit2026} and UppASD\cite{hellsvik2019}.

OpenFerro lets users extend a Hamiltonian by supplying differentiable energy terms as plain Python functions. In FERAM and UppASD, extending the native interaction engines beyond their supplied functional forms requires implementing energy and force or effective-field contributions in compiled code\cite{feram2026,skubic2008,uppasd2026}. For example, a FERAM study modified the code to introduce alloy interactions and customized external fields\cite{mayer2023}. Extending UppASD's dedicated CUDA backend beyond its implemented interaction forms also requires a corresponding GPU implementation\cite{uppasd2026}. Within MULTIBINIT's polynomial representation, additional terms can be specified and fitted through model files\cite{verstraete2025}. In OpenFerro, automatic differentiation generates forces on all coupled fields from the user's JAX energy function, and JAX compiles and distributes these calculations across the GPU mesh. Model extensions therefore leave the simulation and parallelization core unchanged.

The same flexibility extends to enhanced sampling. MULTIBINIT provides hybrid Monte Carlo, while UppASD includes Wang--Landau sampling\cite{verstraete2025,uppasd2026}. OpenFerro supports user-defined biasing potentials: additional energy terms that promote exploration of otherwise rarely visited states. Users express the bias as a differentiable function of the lattice fields or of a chosen system property, such as the overall polarization. Automatic differentiation propagates its derivatives to all participating fields, so changing the bias or its collective variables does not require a separate force implementation.

Earlier polar-model benchmarks also provide context for OpenFerro's scalability. Prokhorenko et al. reported effective-Hamiltonian simulations with long-range dipolar forces for systems of order $10^6$ particles on a single GPU\cite{prokhorenko2018}. Ma et al. demonstrated $128^3$ perovskite cells, representing 10 million atoms, on one CPU core\cite{ma2025}. For atomistic spin dynamics, Etz et al. reported UppASD scalability tests with at least 1.5 billion atomic sites on 13,824 processors of a Cray XE6\cite{etz2015}. Here, OpenFerro advances $840^3=592.7$ million BaTiO$_3$ unit cells in NPT dynamics on one node with four NVIDIA A100 GPUs, retaining local polar modes, acoustic displacements, homogeneous strain, and FFT-based Ewald dipolar interactions (Fig.~\ref{fig:benchmark}). This benchmark represents 3 billion atoms through the retained lattice variables. 

%%%%%%%%%%%%%%%%%%%%%%%%%%%%%%%%%%%%%%%%%%%%%%%%%%%%%%%%%%%%%%%%%%%%%%%%%%%%%%%
\begin{figure*}[t]
  \centering
  \includegraphics[width=\textwidth]{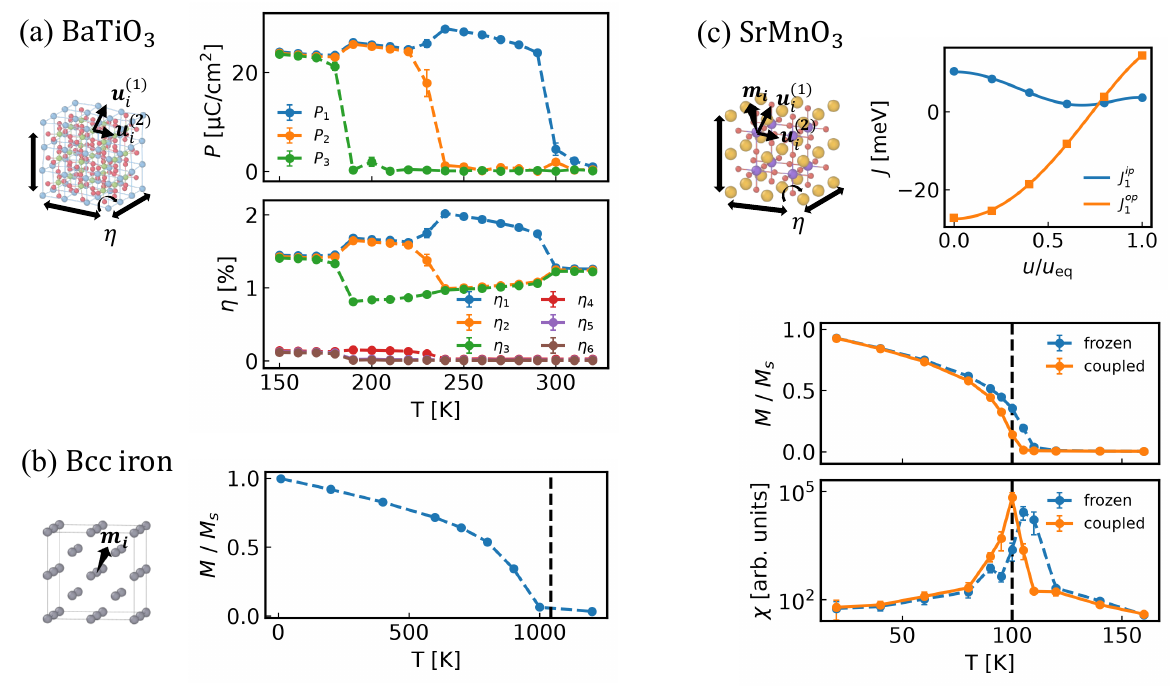}
  \caption{\textbf{Thermodynamics of polar, magnetic, and coupled spin-lattice models.} (a) Spontaneous polarization components $P_1$, $P_2$, and $P_3$ (top) and homogeneous strain components (bottom) versus temperature during cooling of BaTiO$_3$. Polarization components are sorted by magnitude; normal and shear strain components are sorted separately. (b) Normalized magnetization $M/M_s$ versus temperature during heating of the bcc-Fe Heisenberg model. The dashed vertical line marks the experimental Curie temperature, 1043~K. (c) In-plane and out-of-plane first-neighbor exchange couplings, $J_1^{\mathrm{ip}}$ and $J_1^{\mathrm{op}}$, versus normalized soft-mode amplitude $u/u_{\mathrm{eq}}$ (top), and normalized magnetization $M/M_s$ (middle) and magnetic susceptibility $\chi$ (bottom) versus temperature for SrMnO$_3$ at 5\% biaxial tensile strain. The temperature-dependent curves compare frozen-lattice and coupled spin-lattice dynamics in $64^3$ cells. The dashed vertical line marks the approximately 100~K magnetic transition reported for the constrained reference model\cite{edstrom2018}.}
  \label{fig:bto}
\end{figure*}
%%%%%%%%%%%%%%%%%%%%%%%%%%%%%%%%%%%%%%%%%%%%%%%%%%%%%%%%%%%%%%%%%%%%%%%%%%%%%%%

\subsection{General tool for thermodynamics across ferroic orders}
%% This subsection has been finalized. All other sections describing case studies should follow the style of this section. 

We first validate this framework against established thermodynamic results for three distinct models: the standard effective Hamiltonian for BaTiO$_3$, a ferromagnetic bcc-Fe Heisenberg model, and a coupled spin-lattice model for multiferroic SrMnO$_3$.

\subsubsection{Ferroelectric \texorpdfstring{BaTiO$_3$}{BaTiO3}}

We reproduce the standard first-principles effective Hamiltonian of Zhong, Vanderbilt, and Rabe\cite{zhong1995}. Each perovskite cell carries a 3D soft-mode displacement $\mathbf{u}_i$ and a 3D acoustic displacement $\mathbf{v}_i$. $\mathbf{d}_i=Z^{*}\mathbf{u}_i$ defines the local electric dipole moment, where $Z^{*}$ is the effective Born charge of the soft mode.
The homogeneous global strain is a $3\times 3$ symmetric tensor, which can be reduced to a 6-component vector $\bm{\eta}$. The total potential energy of the system is given by
\begin{equation}\label{eq:bto}
  \begin{split}
  E(\mathbf{u}, \mathbf{v}, \bm{\eta}) = & E_{\mathrm{self}}(\mathbf{u}) + E_{\mathrm{dpl}}(\mathbf{u}) + E_{\mathrm{short}}(\mathbf{u}) \\
  & + E_{\mathrm{elas}}(\bm{\eta}, \mathbf{v}) + E_{\mathrm{int}}(\mathbf{u},\mathbf{v},\bm{\eta}).
  \end{split}
\end{equation}
The energy includes on-site potential energy (E$_{\mathrm{self}}$), short-range dipolar interactions (E$_{\mathrm{short}}$), long-range dipolar interactions (as an Ewald sum, E$_{\mathrm{dpl}}$), elastic energy (E$_{\mathrm{elas}}$), and strain-dipolar coupling (E$_{\mathrm{int}}$). The detailed expressions are given in ref.~\citenum{zhong1995}.

We first relax a $24^3$ cell of BaTiO$_3$ with a hydrostatic pressure of $-48$ kbar. This hydrostatic pressure is introduced in ref.~\citenum{zhong1995} to empirically correct for the intrinsic error in the first principles-based model parameters.
Then, we perform an NPT simulation of the relaxed cell with a timestep of 2 fs. The system is cooled from 320 to 150 K. At each temperature, 10 ps of equilibration preceded 50 ps of NPT sampling. Figure~\ref{fig:bto}(a) shows the temperature dependence of the spontaneous polarization (using centrosymmetric cubic structure as the reference of zero polarization) and the global strain components. On cooling, the number of nonzero polarization components changes from zero to one, two, and three. Discontinuities occur between the sampled 300-290 K, 240-220 K, and 190-180 K intervals, resolving the cubic-tetragonal-orthorhombic-rhombohedral sequence. Strain changes in concert with polarization, demonstrating that mutual field interactions and the global-strain integrator act within one force assembly. These results are consistent with the results of the Monte Carlo simulations of ref.~\citenum{zhong1995}.  

It was noted in ref.~\citenum{zhong1995} that removing all terms relevant to $\mathbf{v}$ in the Hamiltonian, i.e. assuming acoustic modes are not coupled to the soft mode,  leads to negligible difference in the phase transition temperatures. We verify this in our simulations using the reduced Hamiltonian:
\begin{equation}
  \begin{split}
  E(\mathbf{u}, \bm{\eta}) = & E_{\mathrm{self}}(\mathbf{u}) + E_{\mathrm{dpl}}(\mathbf{u}) + E_{\mathrm{short}}(\mathbf{u}) \\
  & + E_{\mathrm{elas}}^{\mathrm{h}}(\bm{\eta}) + E_{\mathrm{int}}^{\mathrm{h}}(\mathbf{u},\bm{\eta}).
  \end{split}
  \label{eq:bto_reduced}
\end{equation}
NPT simulations based on this reduced Hamiltonian find no noticeable difference in the temperature dependence of the spontaneous polarization. 

\subsubsection{Ferromagnetic bcc Fe}

We reproduce the four-shell classical Heisenberg model of Tao \textit{et al.}\cite{tao2005} for body-centered-cubic (bcc) iron. Each bcc site carries a spin $\mathbf{m}_i$ with fixed magnitude $M_s=2.23\ \mu_{\mathrm B}$. The energy comprises four isotropic exchange shells. 

We perform an NVT simulation of a $20^3$ bcc lattice of iron with a timestep of 0.2 fs. The cell is heated from 10 to 1200 K. At each temperature, 1 ps of equilibration preceded 2 ps of sampling. Figure~\ref{fig:bto}(b) shows the temperature dependence of the normalized magnetization $M/M_s$, consistent with the results of ref.~\citenum{tao2005}. On heating, the magnetization decreases continuously and loses most of its order between 900 and 1000 K, near the experimental Curie temperature of 1043 K (marked by a vertical line). These results show that the same energy-to-force interface used for flexible lattice-field dynamics also produces stable constrained-spin dynamics.

\subsubsection{Spin-lattice coupling in \texorpdfstring{SrMnO$_3$}{SrMnO3}}

We reproduce the first-principles spin-lattice-coupled model of SrMnO$_3$ from ref.~\citenum{edstrom2018}. That work evolves the lattice at fixed spin, or the spins at fixed lattice, but not both at once. Each cell carries a soft-mode displacement $\mathbf{u}_i$, an acoustic displacement $\mathbf{v}_i$, and a spin $\mathbf{m}_i$ with unit magnitude. The homogeneous strain is represented by a 6-component vector $\bm{\eta}$.
We generalize the model to enable concurrent dynamics of the spin and lattice degrees of freedom in the unified framework of OpenFerro. In our model, a biaxial strain is clamped as $\eta_1=\eta_2=5\%$, and the out-of-plane strain is clamped as $\eta_3=-2.85\%$, for which the system remains in a ferroelectric state with [110] polarization for temperatures up to 200 K. Under such a strain, the spin-lattice coupling can be derived from the first-principles data available from ref.~\citenum{edstrom2018}. The Hamiltonian is given by
\begin{equation}
 H=H_{\mathrm{FE}}^{G}(\bm{u},\bm{v},\bm{\eta})
 +H_{\mathrm{spin}}(\bm{m};\bm{u})
 -H_{\mathrm{spin}}(\bm{m}^{G};\bm{u}).
 \label{eq:smo}
\end{equation}
Here, the ferroelectric Hamiltonian $H_{\mathrm{FE}}^{G}$ is in the standard form of ref.~\citenum{zhong1995} (Eq.~\eqref{eq:bto}), which was parameterized in a G-type antiferromagnetic reference\cite{edstrom2018}.
The magnetic energy $H_{\mathrm{spin}}(\bm{m};\bm{u})$ is a classical Heisenberg Hamiltonian,
\begin{equation}
  H_{\mathrm{spin}}
  =-\sum_{\langle ij\rangle_1}J_{1,ij}(s_{ij})\mathbf{m}_i\cdot\mathbf{m}_j
  -\sum_{\langle ij\rangle_2}J_{2,ij}\mathbf{m}_i\cdot\mathbf{m}_j.
  \label{eq:hspin}
\end{equation}
Here, $J_{1,ij}(s_{ij})$ and $J_{2,ij}$ are the first-neighbor and second-neighbor exchange couplings, respectively. First-neighbor couplings are tetragonal and depend on the local soft-mode displacement, where the bond coordinate is defined as $s_{ij}=(|\mathbf{u}_i|^2+|\mathbf{u}_j|^2)/(2u_{\mathrm{eq}}^2)$ with $u_{\mathrm{eq}}=0.23$ \AA. In-plane and out-of-plane first-neighbor bonds use distinct $J_1^{\mathrm{ip}}(s_{ij})$ and $J_1^{\mathrm{op}}(s_{ij})$, which are parameterized as cubic polynomials fitted to the published homogeneous [110] $J_1(u/u_{\mathrm{eq}})$ curve. The first panel of Fig.~\ref{fig:bto}(c) compares the fit with the reference points.  
Second-neighbor couplings are held at the relaxed ferroelectric values $J_2^{[110]}=0.186$~meV, $J_2^{[1\bar{1}0]}=0.779$~meV, and $J_2^{\mathrm{op}}=0.062$~meV. 
The model described by Eq.~\eqref{eq:smo} reduces exactly to the published structural Hamiltonian with the spins clamped in the G-type antiferromagnetic reference configuration ($\mathbf{m}^{G}$)\cite{edstrom2018}, while deviation from the reference spin configuration generates forces that drive spin and lattice degrees of freedom concurrently.

%%%%%%%%%%%%%%%%%%%%%%%%%%%%%%%%%%%%%%%%%%%%%%%%%%%%%%%%%%%%%%%%%%%%%%%%%%%%%%%
\begin{figure*}[!t]
  \centering
  \includegraphics[width=0.95\textwidth]{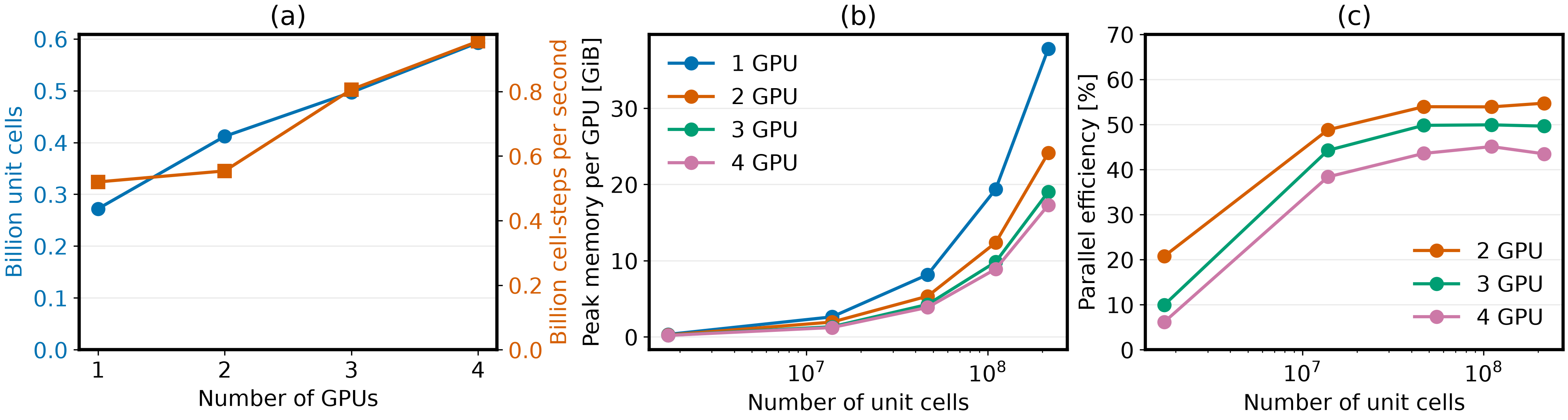}
  \caption{\textbf{Throughput and single-node scaling of BaTiO$_3$ NPT dynamics.} (a) Largest timed systems found by the memory-constrained size search (blue, left axis) and throughput at those sizes (orange, right axis). One cell-step advances all retained fields of one unit cell by a complete integration step. (b) Peak allocator memory per GPU and (c) parallel efficiency for $L=120$, 240, 360, 480, and 600. Both panels use a logarithmic system-size axis. Parallel efficiency is the single-GPU step time divided by the GPU count and parallel step time. Timings are medians of three 100-step measurements after compilation and warmup, excluding trajectory output.}
  \label{fig:benchmark}
\end{figure*}
%%%%%%%%%%%%%%%%%%%%%%%%%%%%%%%%%%%%%%%%%%%%%%%%%%%%%%%%%%%%%%%%%%%%%%%%%%%%%%%

We perform two NVT simulations on a $64^3$ cell with a timestep of 0.2~fs, heating from 20 to 160~K, where the system transforms from a ferroelectric, ferromagnetic state to a ferroelectric, paramagnetic state.
In the first simulation (the frozen case), the lattice degrees of freedom are frozen in the relaxed ferroelectric state with [110] polarization, while the spins are allowed to evolve. This simulation reproduces the results of the spin-only simulations performed in ref.~\citenum{edstrom2018}. In the second simulation (the coupled case), the lattice degrees of freedom are allowed to evolve concurrently with the spins. This simulation is used to demonstrate the generality of the OpenFerro framework for modelling multiferroic materials.

At each temperature, 1~ps of equilibration preceded 10~ps of sampling. The second and third panels of Fig.~\ref{fig:bto}(c) show the magnetization magnitude and the magnetic susceptibility. On heating, both calculations lose ferromagnetic order near 100~K, which is consistent with the magnetic transition reported for the constrained reference model\cite{edstrom2018}. In our coupled simulation, the average magnitude of local mode, $\langle |\mathbf{u}_i| \rangle$, drops from $0.24\,\text{\AA}$ (at 20 K) to $0.22\,\text{\AA}$ (at 160 K), indicating only a minor deviation from the reference value $u_{\mathrm{eq}}=0.23\,\text{\AA}$ used to parameterize the exchange couplings and therefore validating the applicability of the model in this temperature range.

\subsection{Overcoming the length scale: superlattice and nanoparticles}

We next demonstrate mesoscopic lattice dynamics with millions to hundreds of millions of perovskite cells on a single GPU node. We first benchmark the accessible system sizes using the BaTiO$_3$ Hamiltonian, then examine polarization textures in PbTiO$_3$/SrTiO$_3$ superlattices and BaTiO$_3$ nanoparticles.

\subsubsection{Parallelization enables billion-atom-level simulations}

We benchmark OpenFerro's scalability through NPT simulations of BaTiO$_3$ using the complete Hamiltonian of ref.~\citenum{zhong1995} (Eq.~\eqref{eq:bto}). Simulations use a 2~fs timestep on one to four NVIDIA A100-SXM4-80GB GPUs within a single GPU node on the Perlmutter supercomputer, with memory capped at 60~GiB per GPU.

We measure scalability in terms of capacity and throughput. Capacity is the largest cubic simulation cell that fits within the memory cap and completes a timed run. Throughput, $N/t_{\mathrm{step}}$, is the number of cell-steps per wall-clock second, where one cell-step advances all retained fields of one unit cell through a complete integration step. The system size that maximizes throughput need not maximize capacity.

Figure~\ref{fig:benchmark}(a) shows the capacity and corresponding throughput. Capacity increases from $648^3=272.1$ million unit cells on one GPU to $840^3=592.7$ million on four, while throughput increases from 0.520 to 0.955 billion cell-steps per second. The largest system spans 331.6 nm along each axis and represents 3 billion atoms through the retained lattice fields. A complete integration step takes 0.620 s. At the smaller size of $480^3=110.6$ million cells (not shown in panel (a)), four GPUs reach 1.00 billion cell-steps per second. This higher throughput for smaller systems arises from the FFT-based Ewald cost, which scales as $N\log N$. Extrapolating the kernel timing yields 1.57 ns of simulated dynamics per day for the 110.6-million-cell system.

\begin{figure*}[!t]
  \centering
  \includegraphics[width=\textwidth]{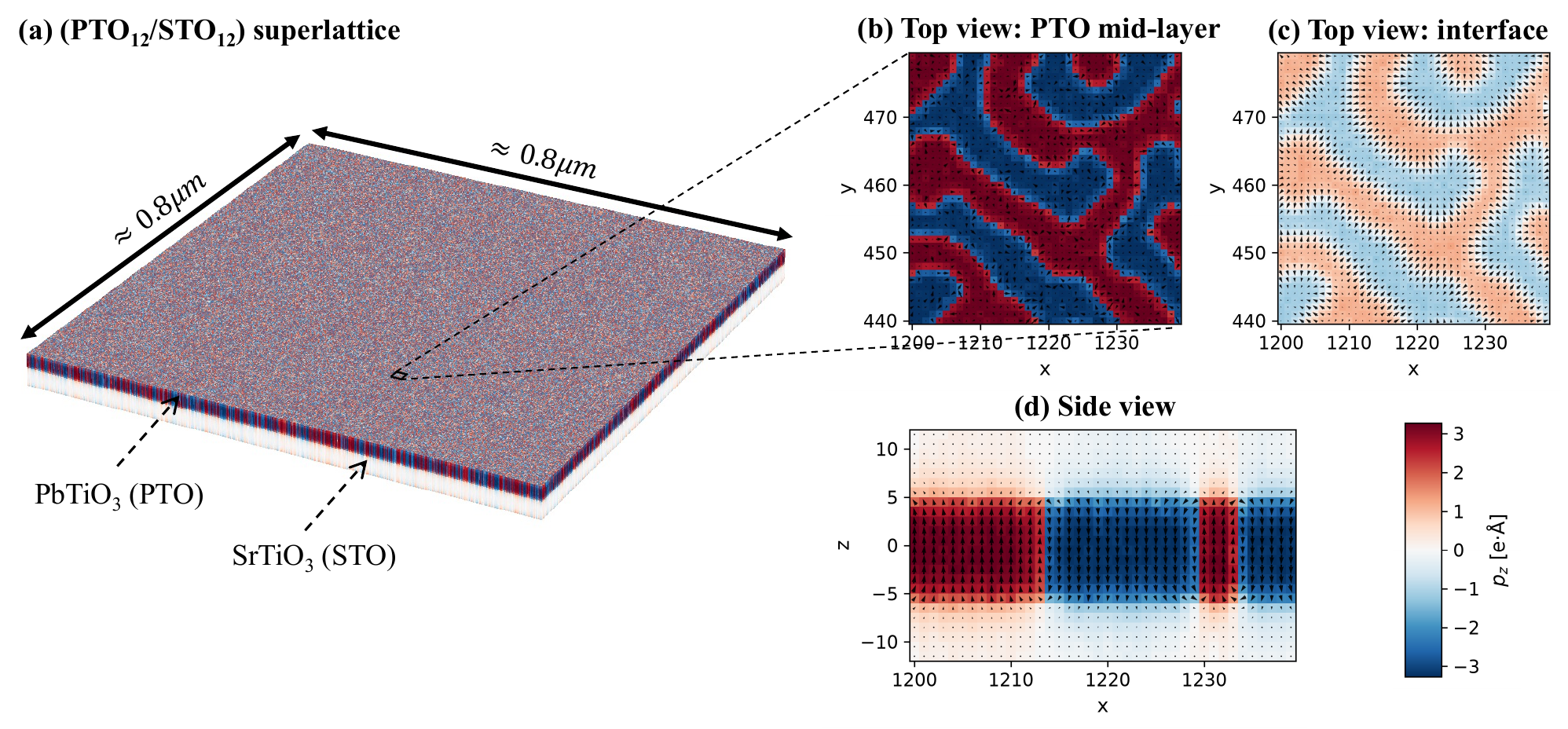}
  \caption{\textbf{Polarization texture in a PbTiO$_3$/SrTiO$_3$ superlattice.} (a) $(\mathrm{PbTiO}_3)_{12}/(\mathrm{SrTiO}_3)_{12}$ superlattice after cooling to 10~K. The slab is $2000\times2000$ cells laterally, about $0.8~\mu\mathrm{m}$ on a side, and 24 cells thick. PbTiO$_3$ forms the polar layers and SrTiO$_3$ the weakly polarized layers. Dashed lines mark the window enlarged in (b)-(d). (b) Out-of-plane local dipole component $p_z$ in the middle of the PbTiO$_3$ block, inside a $40\times40$ window. The pattern is a labyrinth of 180$^\circ$ domains, and the arrows are the in-plane local dipoles at each cell. (c) The same window at the PbTiO$_3$/SrTiO$_3$ interface, where $|p_z|$ is smaller. (d) The $x$-$z$ cut through the center of the window. The $z$ coordinate is shifted so the PbTiO$_3$ layers are centered and SrTiO$_3$ lies equally above and below. Arrows are the projection of the local dipole onto this plane. Color in (b)-(d) is the time-averaged $p_z$ in $e\,\text{\AA}$, and coordinates are in lattice-cell units.}
  \label{fig:ptosto}
\end{figure*}

Parallelization also reduces the memory required on each GPU [Fig.~\ref{fig:benchmark}(b)]. For $600^3=216$ million cells, peak memory per GPU falls from 37.82 GiB on one GPU to 17.30 GiB on four. Parallel efficiency increases with system size as communication overhead is spread over more local work [Fig.~\ref{fig:benchmark}(c)]. Four-GPU efficiency reaches approximately 43-45\% for the larger benchmarks, compared with 6\% for systems of roughly $10^6$ unit cells. Together, these gains in capacity and throughput allow cell-scale polar and elastic structure to remain explicit across hundreds of nanometres on a single GPU node.

The demonstrated capacity suggests that a thin-film geometry, rather than a cube, can extend the lateral dimensions to micrometre scale. This motivates the PbTiO$_3$/SrTiO$_3$ superlattice considered next.

\subsubsection{PTO-STO superlattices}

We construct an effective Hamiltonian for a PbTiO$_3$/SrTiO$_3$ superlattice from the first-principles models of refs.~\citenum{paul2017,nishimatsu2016}.
The superlattice is a periodic stack of $m$ PbTiO$_3$ (PTO) layers and $n$ SrTiO$_3$ (STO) layers along $z$. The simulation supercell therefore consists of an $L\times L\times(m+n)$ lattice, where $L$ is the number of lattice sites along the lateral direction.
Each lattice site carries a 3D local soft-mode displacement $\mathbf{u}_i$.
The homogeneous strain of the PTO block is a 6-component vector $\bm{\eta}$. We assume the STO block to have zero strain. 
Consequently, the in-plane strains $\eta_{1}=\eta_{2}$ are fixed such that the PTO in-plane lattice constant matches the equilibrium lattice constant of STO. This applies a compressive strain to the PTO block and favors out-of-plane polarization. The out-of-plane strain $\bm{\eta}_{3}$ is free. Shear strains and inhomogeneous strains are omitted for simplicity.

The potential energy of the superlattice has the same decomposition as the reduced ferroelectric Hamiltonian in Eq.~\eqref{eq:bto_reduced},
\begin{equation}
\begin{aligned}
E(\mathbf{u},\bm{\eta})
&= E_{\mathrm{self}}(\mathbf{u}) + E_{\mathrm{short}}(\mathbf{u}) + E_{\mathrm{dpl}}(\mathbf{u}) \\
&\quad + E_{\mathrm{elas}}^{\mathrm{h}}(\bm{\eta}) + E_{\mathrm{int}}^{\mathrm{h}}(\mathbf{u},\bm{\eta}).
\end{aligned}
\label{eq:ptosto}
\end{equation}
In the interior of each block, the onsite ($E_{\mathrm{self}}$), short-range ($E_{\mathrm{short}}$), and strain-dipole ($E_{\mathrm{int}}^{\mathrm{h}}$) parameters are those of PbTiO$_3$\cite{paul2017} or SrTiO$_3$\cite{nishimatsu2016}.
At the two interfaces, each of these parameters is approximated as an arithmetic mean of the two constituents.
For the long-range dipolar interactions ($E_{\mathrm{dpl}}$), the local dipole moment is calculated as $\mathbf{p}_i = Z_i^{*} \mathbf{u}_i$, where $Z_i^{*}$ is the effective Born charge of site $i$. The background dielectric constant is taken as PbTiO$_3$'s background dielectric constant. The elastic energy is only calculated for the PTO block because the STO block is assumed to have zero strain. For isobaric simulations, the pressure is applied to the PTO block only.

Here, we simulate the PTO-STO superlattice with $m=n=12$ and a lateral size of $L=2000$.
The resulting $2000\times2000\times24$ simulation box is equivalent to an atomic system of 0.48~billion atoms. 
We cool the superlattice in the NPT ensemble from 300 to 10~K in steps of 10~K at zero electric field, holding for 50~ps at each temperature. After the cooling, we perform a further 10~ps of NPT sampling at 10~K. The local dipole moments ($\mathbf{p}_i$) are sampled every 1~ps and then averaged over this period to yield the quenched polarization texture (Fig.~\ref{fig:ptosto}).  This simulation uses a 2~fs timestep and four NVIDIA A100 GPUs.

Fig.~\ref{fig:ptosto}(a) shows that the PTO block forms a labyrinth of 180$^\circ$ domains. Fig.~\ref{fig:ptosto}(b,c,d) are zoom-in views of a small $40\times40\times24$ window marked in Fig.~\ref{fig:ptosto}(a). The color of the pixels indicates the $z$-component of the local dipole $\mathbf{p}_i$. The arrows are the projection of the local dipole onto each displayed plane.
Fig.~\ref{fig:ptosto}(b) shows that, in the middle $x$-$y$ layer of the PTO block, the dipole is predominantly out of plane. Fig.~\ref{fig:ptosto}(c) provides a close-up view of the PTO/STO interface, where $|p_z|$ is smaller and the in-plane dipole lies perpendicular to the domain walls. This is the feature of the flux-closure-type polarization rotation\cite{zatterin2024}. The flux-closure pattern is also shown by Fig.~\ref{fig:ptosto}(d), which displays the $x$-$z$ cut through the center of the window. The $z$ coordinate is shifted so the PTO layers are centered and STO lies equally above and below. Here, the local dipole stays out of plane through the interior of the PTO block and turns toward the layer plane at the walls and near both interfaces. This is a flux-closure, or N\'eel, rotation.

These observations agree with the phase-field ground-state simulation of PTO/STO superlattices in ref.~\citenum{zatterin2024}. This type of mesoscale polarization texture study is commonly studied using phase-field models. Here, we have demonstrated that, with the unified OpenFerro framework and minimal implementation effort, one can simulate this type of nanostructure on a $\mu$m length scale with atomistic resolution without relying on a continuum model.

\subsubsection{\texorpdfstring{BaTiO$_3$ nanoparticles}{BaTiO3 nanoparticles}}

We construct an effective Hamiltonian for a BaTiO$_3$ nanoparticle. 
The nanoparticle is a spherical region of diameter $D$ (number of lattice sites along the longest line segment) embedded in a cubic, periodic supercell consisting of a $L\times L\times L$ lattice, where each lattice site carries a 3D soft-mode displacement $\mathbf{u}_i$. For sites inside the sphere, $\mathbf{u}_i$ is a free variable. For sites outside the sphere, $\mathbf{u}_i$ is held at $\mathbf{0}$ to represent a vacuum environment.
The homogeneous strain of the nanoparticle is represented by a 6-component vector $\bm{\eta}$. Inhomogeneous strain is omitted.

We construct the nanoparticle energy from Eq.~\eqref{eq:bto_reduced} by adding two additional terms ($E_{\mathrm{surf}}$ and $E_{\mathrm{shell}}$):
\begin{equation}
\begin{aligned}
E(\mathbf{u},\bm{\eta})
& = E_{\mathrm{self}}(\mathbf{u}) + E_{\mathrm{short}}(\mathbf{u}) + E_{\mathrm{dpl}}(\mathbf{u}) + E_{\mathrm{elas}}^{\mathrm{h}}(\bm{\eta})\\
& \quad + E_{\mathrm{int}}^{\mathrm{h}}(\mathbf{u},\bm{\eta}) + E_{\mathrm{surf}}(\mathbf{u}) + E_{\mathrm{shell}}(\mathbf{u}).
\end{aligned}
\label{eq:bto_np}
\end{equation}
Because $\mathbf{u}_i$ is zero outside the sphere, the short-range interaction ($E_{\mathrm{short}}$) and strain-dipolar coupling ($E_{\mathrm{int}}^{\mathrm{h}}$) are strictly restricted to the nanoparticle. However, the Ewald dipolar interactions ($E_{\mathrm{dpl}}$) still act between periodic images of the nanoparticle, which is spurious for describing a nanoparticle in vacuum.
We impose a spherical-vacuum boundary condition on the Ewald sum by adding $E_{\mathrm{surf}}$\cite{ballenegger2014}, given as
\begin{equation}
  E_{\mathrm{surf}}
  = \frac{Z^{*2}}{\epsilon_{\infty}}
  \frac{\bigl|\sum_i \mathbf{u}_i\bigr|^2}{6\epsilon_0 V_{\mathrm{box}}}.
  \label{eq:bto_surf}
\end{equation}
Here, $\epsilon_{\infty}$ is the same background dielectric constant as that used in $E_{\mathrm{dpl}}$. $V_{\mathrm{box}}$ is the volume of the cubic supercell. With the net dipole $\mathbf{M}=(Z^{*}/\sqrt{\epsilon_{\infty}})\sum_i\mathbf{u}_i$, this is the depolarization energy $|\mathbf{M}|^2/(6\epsilon_0 V_{\mathrm{box}})$ of a spherical sample of that volume surrounded by vacuum. An isolated-particle interpretation requires increasing $L$ at fixed $D$ until the corrected electrostatic energy and relevant observables converge.

The second additional term, $E_{\mathrm{shell}}$, corrects the vacuum-particle interfacial energy by adding onsite quadratic energy in outer shell sites close to the interface. Let $i\in\mathrm{shell}$ be the sites within $\delta R$ cells from the vacuum-particle interface. $E_{\mathrm{shell}}$ is given by
\begin{equation}
E_{\mathrm{shell}} = \Delta\kappa_2 \sum_{i\in\mathrm{shell}} |\mathbf{u}_i|^2,
\label{eq:bto_shell}
\end{equation}
$\Delta\kappa_2$ is a positive constant and $E_{\mathrm{shell}}$ weakens the soft-mode displacement of the shell sites.  $E_{\mathrm{shell}}$ is not derived from first principles. In our simulations, we use empirical values of $\Delta\kappa_2=5.502~\mathrm{eV}\,\text{\AA}^{-2}$ and $\delta R=8$ for the BaTiO$_3$ nanoparticle. 

Because of the empirical nature of the constructed interfacial energy, the nanoparticle model here represents a minimal, semi-quantitative model for a BaTiO$_3$ nanoparticle. A more rigorous model would derive an interfacial energy involving modifications to all bulk parameters from first principles. In this work, we focus on demonstrating the capability of OpenFerro to simulate this type of nanostructure on sub-micrometer length scales, while leaving more sophisticated models for future work.

\begin{figure*}[!t]
  \centering
  \includegraphics[width=\textwidth]{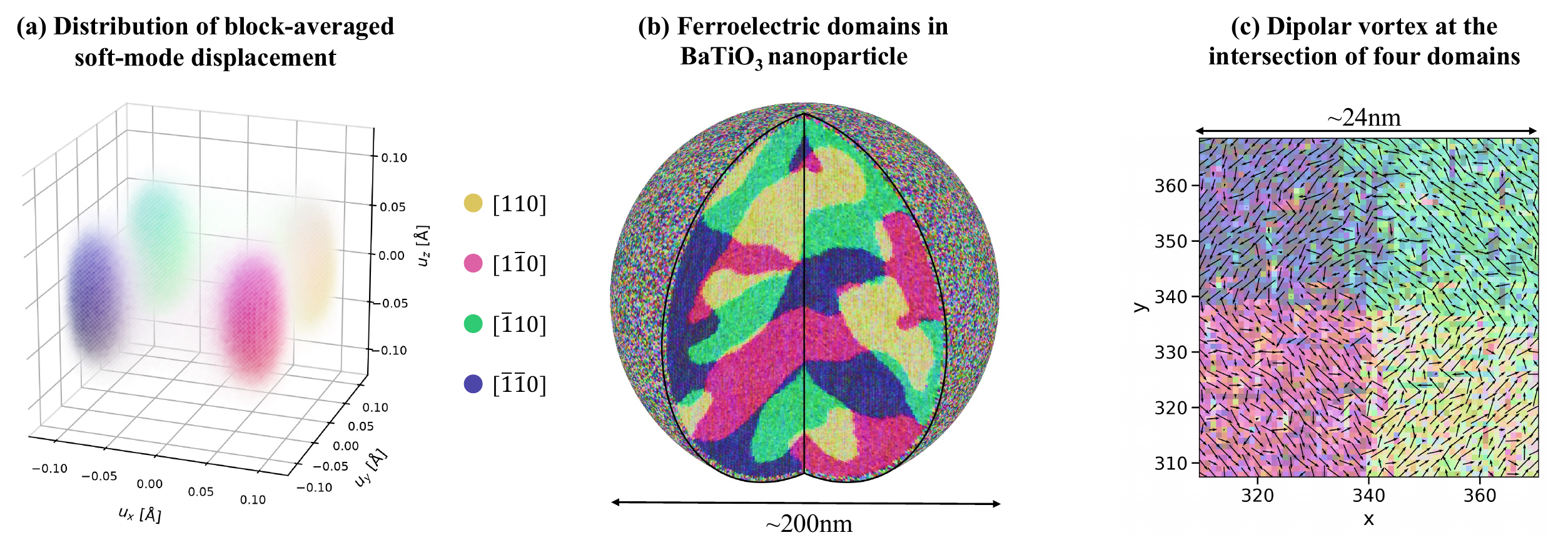}
  \caption{\textbf{Polarization domains and a dipolar vortex in a BaTiO$_3$ nanoparticle.} (a)~Distribution of block-averaged soft-mode displacements in the core of the $D=502$ BaTiO$_3$ nanoparticle (about 200~nm in diameter) at the end of the 250~K trajectory. The core excludes the outer shell of thickness $\delta R=8$ cells, and each displacement is the mean over a $3\times 3\times 3$ block. Colors encode displacement direction through the RGB mapping \((R,G,B)=(\mathbf{1}+\hat{\mathbf u})/2\): yellow, pink, green, and blue mark $[110]$, $[1\bar{1}0]$, $[\bar{1}10]$, and $[\bar{1}\bar{1}0]$. Saturation and opacity increase linearly with the bin count and reach full strength at half the peak count. The four lobes are these in-plane $\langle 110\rangle$ variants, with a spread along $u_z$. (b)~Quarter cutaway of the same configuration, opened along $+x$ between the $(110)$ and $(1\bar{1}0)$ planes and colored by the same map after the same block average. The interior is divided into extended domains of the four variants, and the outer shell remains finely mixed. (c)~Unaveraged soft-mode displacement on the plane $z=214$, in a $61\times 61$ window (about 24~nm on a side) centered where the four domains meet. Arrows are $(u_x,u_y)$ at every second cell, and the background uses the same directional coloring as (b). The displacement circulates counterclockwise about the junction. Coordinates in (c) are lattice-cell indices.}
  \label{fig:nanoparticle} 
\end{figure*}

Here, we simulate a BaTiO$_3$ nanoparticle with $D=502$, corresponding to a diameter of 200 nm. The enclosing box ($L=512$) has $512^3$ sites; only sites inside the nanoparticle sphere contribute to the particle volume and homogeneous elastic energy. Starting from small random local modes, we simulate NPT dynamics of the nanoparticle with temperature cooled from 320 to 250 K in 10 K intervals at $-48$ kbar, with 25 ps of NPT equilibration at each temperature. At 250 K, we continue NPT sampling for 1 ns. The simulation uses a 2 fs timestep and four NVIDIA A100 GPUs. 

At 250~K, bulk BaTiO$_3$ described by the bulk effective Hamiltonian (Eq.~\eqref{eq:bto_reduced}) is tetragonal, with a uniform [001] polarization (Fig.~\ref{fig:bto}) and an average soft-mode displacement of the order of $0.1$~\AA. The nanoparticle responds differently due to the strong depolarizing effect of the surrounding vacuum. The net spontaneous polarization stays negligible, but ferroelectric domains form. Through the 1 ns NPT simulation at 250~K, the ferroelectric domains display slow coarsening. 
Here, we analyze the slowly evolving domain structure using a representative snapshot of the dipole configuration captured at the end of the 1~ns simulation.
Figure~\ref{fig:nanoparticle}(a) shows the distribution of block-averaged soft-mode displacements in the core, which excludes the outer shell of thickness $\delta R=8$ cells. The average is taken over a local $3\times 3\times 3$ window at each site to suppress thermal noise. The weight lies in four lobes along $[110]$, $[1\bar{1}0]$, $[\bar{1}10]$, and $[\bar{1}\bar{1}0]$, with only a small spread along $u_z$. These four in-plane $\langle 110\rangle$ variants are the polar domains of the nanoparticle.

Figure~\ref{fig:nanoparticle}(b) shows their arrangement. A quarter of the sphere is removed along $+x$, between the $(110)$ and $(1\bar{1}0)$ planes. Across the roughly 200~nm diameter, the interior is partitioned into extended domains of the four variants, while the outer shell, where $E_{\mathrm{shell}}$ weakens the soft mode, stays finely mixed. Where the four variants of polar domains meet, the soft-mode displacement (without block averaging) circulates. Figure~\ref{fig:nanoparticle}(c) shows one such junction, in a $61\times 61$ window of about 24~nm on a side. The arrows are the in-plane displacement $(u_x,u_y)$ of individual cells. $[\bar{1}\bar{1}0]$ and $[\bar{1}10]$ occupy the upper half of the window, and $[1\bar{1}0]$ and $[110]$ occupy the lower half, so the displacement circulates counterclockwise about the intersection. The four domains therefore meet in a dipolar vortex.

Polarization vortices in ferroelectric nanoparticles have been reported in effective-Hamiltonian simulations\cite{fu2003,naumov2004,naumov2008} and observed experimentally in BaTiO$_3$\cite{karpov2017,jeong2024}. Our 200~nm simulation shows that OpenFerro can follow domain evolution while resolving individual dipoles around the vortices (Fig.~\ref{fig:nanoparticle}).
A more detailed study of the finite-size physics of the nanoparticle will be reported in a later publication~\cite{xie2027nanoparticle}.

\subsection{Overcoming the time scale: rare events and enhanced sampling}
The last section demonstrates that OpenFerro can simulate mesoscopic nanostructures with moderate GPU resources, but a direct simulation still covers only nanoseconds. The timestep is set by the fastest motion retained in the lattice dynamics, so many rare events associated with phase transitions remain inaccessible. Nucleation of a reversed ferroelectric or magnetic domain, creep of a pinned domain wall, collective freezing of polar nanoregions, and diffusion of topological defects are examples. These processes determine coercivity, phase coexistence, and slow relaxation. Their timescales run from nanoseconds to milliseconds or longer, and an unbiased lattice-dynamics simulation therefore remains in the stable or metastable basin where it started. These time-scale limitations can often be addressed by enhanced sampling methods. A large class of such methods biases the Hamiltonian through a few collective variables (CVs), changing how often the rare events are visited\cite{valsson2016}. Another class of enhanced sampling methods, represented by the replica exchange algorithm\cite{swendsen1986,sugita1999}, does not rely on collective variables, but instead constructs multiple ensembles of the same system. At this point, the development of OpenFerro focuses on CV-based methods for reasons of efficiency and throughput.

\subsubsection{Enhanced sampling with biasing potential}
A CV $\bm{s}(\bm{x})$ is typically chosen in a way that its values separate the metastable states and the paths that connect them. For enhanced sampling with the CV, a biasing potential $V_{\mathrm{b}}(\bm{s})$ is added to the Hamiltonian, and a canonical-ensemble simulation will sample the Boltzmann weight of $H+V_{\mathrm{b}}$. In the current context, $\bm{x}$ includes all the lattice fields of the system with or without the global strain tensor. 

Under the biasing potential, the equilibrium free energy of the original Hamiltonian along $\bm{s}$ is connected to the biased Boltzmann distribution ($P_{\mathrm{b}}(\bm{s})$) by
\begin{equation}
F(\bm{s})=-k_{\mathrm{B}}T\ln P_{\mathrm{b}}(\bm{s})-V_{\mathrm{b}}(\bm{s})+\mathrm{const},
\label{eq:reweight}
\end{equation}
where $P_{\mathrm{b}}(\bm{s})$ is calculated as the histogram of $\bm{s}$ in the biased ensemble. Umbrella sampling\cite{torrie1977}, metadynamics\cite{laio2002}, and variational enhanced sampling\cite{valsson2014} are free-energy calculation methods based on different constructions of the biasing potential. At this point, OpenFerro supports umbrella sampling and metadynamics.

In umbrella sampling, the biasing potential is specified in advance. A restraint, usually harmonic, holds $\bm{s}$ inside one window. Overlapping windows are combined to reconstruct $F(\bm{s})$ across the range of interest. In OpenFerro, such a biasing potential can be directly added to the effective Hamiltonian as a scalar function of the lattice fields and requires no additional modules. So demonstration of the umbrella sampling is omitted here.

In a metadynamics simulation, the biasing potential is not specified in advance, but is built during the simulation and is therefore a function of time, $V_{\mathrm{b}}(\bm{s},t)$. It grows where $\bm{s}$ has already been visited, filling the free-energy wells and driving the system over the barriers between them. As $t\to\infty$, the bias converges asymptotically to a linear function of the free-energy surface,
\begin{equation}
V_{\mathrm{b}}(\bm{s},t\to\infty)=-\alpha F(\bm{s})+C,
\label{eq:meta_limit}
\end{equation}
where $C$ is a constant. In the original formulation of metadynamics, $\alpha=1$\cite{laio2002}. In the well-tempered metadynamics introduced later\cite{barducci2008}, $\alpha=\Delta T/(T+\Delta T)$, with $\Delta T$ a chosen temperature parameter.  

The free-energy surface is thus obtained from the simulation itself, without an \textit{a priori} ansatz such as a Landau polynomial. Earlier studies reconstructed ferroelectric free energies by fitting a presumed analytic form\cite{iniguez2001}, or by thermodynamic integration along selected polarization paths\cite{geneste2009,kumar2010}. An analytic ansatz requires choosing the relevant polynomial terms and truncation order in advance, making it most readily applicable to well-studied materials with an established phenomenological description. Extending it to less familiar materials requires additional knowledge of their phases and couplings. Direct multidimensional sampling by metadynamics reduces this dependence on prior knowledge of the free-energy landscape and provides a general procedure across different materials\cite{xie2026fes}.

OpenFerro provides a metadynamics implementation based on the algorithm of ref.~\citenum{laio2002} and on the conventions of the PLUMED package\cite{bonomi2009plumed, tribello2014plumed}.

\subsubsection{\texorpdfstring{Exploring the polarization free-energy surface of BaTiO$_3$}{Exploring the polarization free-energy surface of BaTiO3}}

\begin{figure*}[!t]
  \centering
  \includegraphics[width=\textwidth]{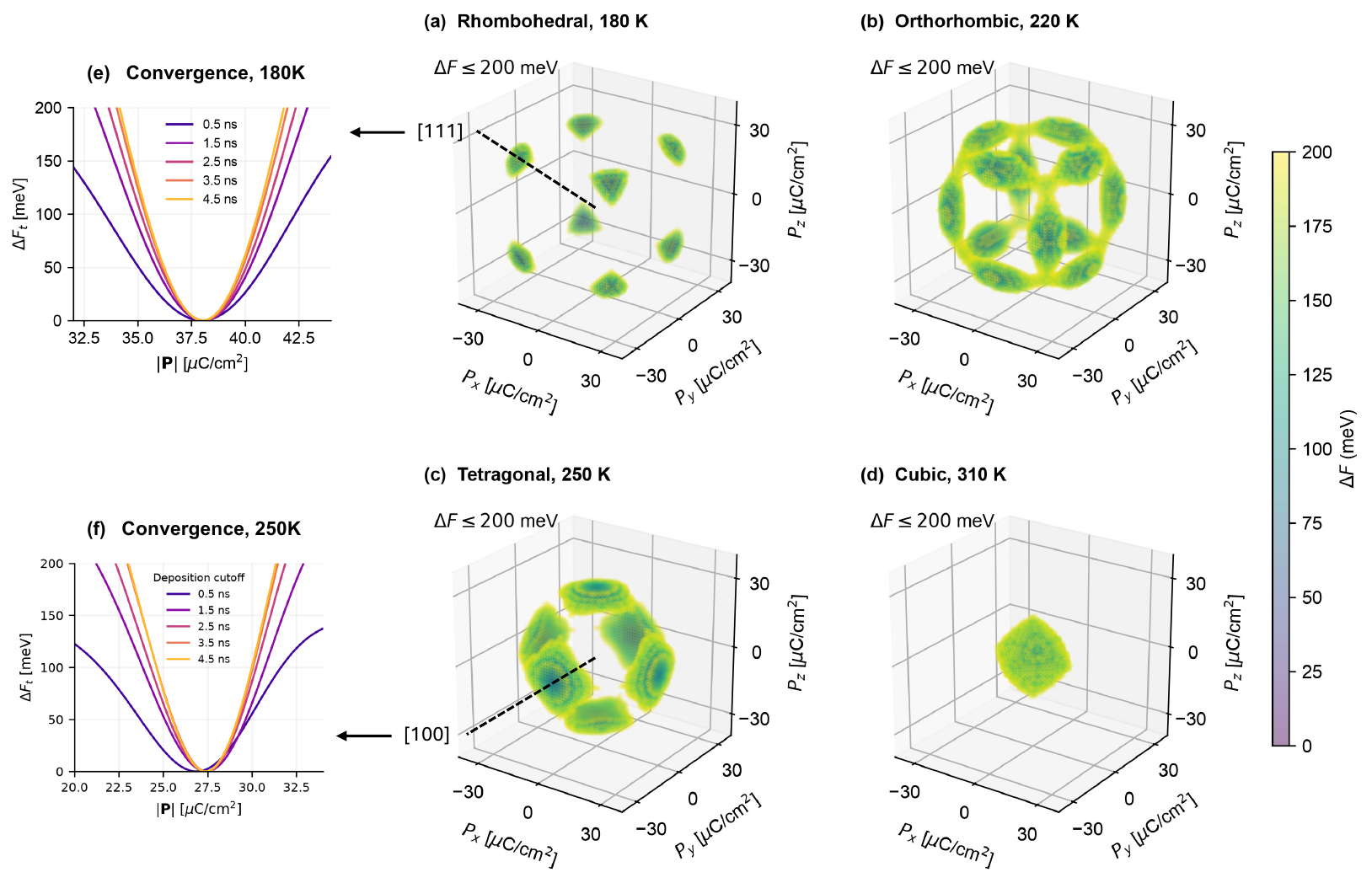}
  \caption{\textbf{Polarization free-energy landscapes of a BaTiO$_3$ cell, reconstructed from the time-dependent metadynamics bias.} (a-d) Symmetry-recovered low-energy regions at 180, 220, 250, and 310~K, respectively. Points satisfy $\Delta F=\widehat F-\min\widehat F\leq200$~meV and are colored by $\Delta F$. (e,f) Directional cuts of the free-energy surface along $[111]$ at 180~K and $[100]$ at 250~K, reconstructed from the bias at the indicated times to examine convergence. Each curve is shifted to its own minimum. Energies refer to the entire simulation cell.}
  \label{fig:metad}
\end{figure*}

For a proper ferroelectric, the three-dimensional macroscopic polarization is a function of the lattice fields in the effective Hamiltonian. In the model of Eq.~\eqref{eq:bto_reduced}, the polarization can be defined as the sum of local dipole moments divided by the volume of the cell, given as
\begin{equation}
\mathbf{P}(\mathbf{u},\bm{\eta})=\frac{Z^{*}}{V(\bm{\eta})}\sum_i\mathbf{u}_i.
\label{eq:polarization}
\end{equation}
Here, $V(\bm{\eta})$ is the volume of the strained cell. 
A rigorous definition of the polarization would require specifying a branch modulo a polarization quantum, which is omitted here for simplicity.

For the NPT simulations considered here, the polarization free-energy surface is the constant-pressure marginal of the Hamiltonian, given as
\begin{equation}
\begin{split}
F(T,\mathbf{P})={}&-k_{\mathrm{B}}T\ln\int\mathcal{D}\mathbf{u}\,\mathrm{d}\bm{\eta}\;
\delta^{(3)}\!\bigl(\mathbf{P}-\mathbf{P}(\mathbf{u},\bm{\eta})\bigr)
\\
&\qquad\times e^{-[E(\mathbf{u},\bm{\eta})+pV(\bm{\eta})]/k_{\mathrm{B}}T}+\text{const},
\end{split}
\label{eq:fes}
\end{equation}
where $p$ is the imposed hydrostatic pressure. Both the local soft modes and the homogeneous strain are integrated over, so this surface includes their thermal fluctuations and strain relaxation at fixed polarization.

We perform independent NPT metadynamics simulations of a $12^3$ cell using the reduced Hamiltonian of Eq.~\eqref{eq:bto_reduced}, at $p=-48$~kbar and temperatures of 180, 220, 250, and 310~K. These temperatures represent the rhombohedral, orthorhombic, tetragonal, and cubic phases found in the cooling simulations. Each trajectory provides approximately 5~ns of sampling.

To avoid repeatedly filling symmetry-equivalent regions, we follow the symmetry-restricted sampling strategy introduced by ref.~\citenum{xie2026fes} and use the three-dimensional CV $\bm{s}=(s_1,s_2,s_3)$ defined as
\begin{equation}
s_1=P_x-P_y,\qquad s_2=P_y-P_z,\qquad s_3=P_z.
\label{eq:meta_chamber}
\end{equation}
The sector $s_i\geq0$ corresponds to $P_x\geq P_y\geq P_z\geq0$. We convert the time-dependent biasing potential from the metadynamics simulation into an estimate of the polarization free-energy surface, $\widehat F_t(\mathbf P)=-V_{\mathrm{b}}(\bm{s}(\mathbf P),t)+C_t$, following Eq.~\eqref{eq:meta_limit}. $C_t$ is a constant shift of the free-energy surface not depending on the polarization.
Because the Hamiltonian and hydrostatic pressure retain cubic symmetry, the surface in the other sectors follows by signed permutations of $(P_x,P_y,P_z)$.

Figure~\ref{fig:metad}(a-d) shows the symmetry-recovered free-energy surface within 200~meV of the minimum at each temperature. At 180~K, the minima lie along the eight $\langle111\rangle$ directions, with a polarization magnitude of approximately $38~\mu\mathrm{C / cm^{2}}$. At 220~K, the low-energy regions are centered on the twelve $\langle110\rangle$ directions, with a magnitude of approximately $33~\mu\mathrm{C / cm^{2}}$. At 250~K, six $\langle100\rangle$ minima remain, with a magnitude of approximately $27~\mu\mathrm{C / cm^{2}}$. At 310~K, the minimum moves to $\mathbf P=0$, consistent with the cubic paraelectric phase.

Figures~\ref{fig:metad}(e,f) examine directional cuts of the free-energy surface along $[111]$ at 180~K and $[100]$ at 250~K, reconstructed from the biasing potential at 0.5, 1.5, 2.5, 3.5, and 4.5~ns. Each curve is shifted to its own minimum to remove the arbitrary additive constant. The minimum positions stabilize early, and the late-time profiles approach a consistent shape in the well-explored low-free-energy region. Comparing these time-dependent reconstructions allows us to examine the asymptotic convergence of the bias shape toward the negative of the formal polarization free-energy surface (Eq.~\eqref{eq:fes}) within that region.

\section{Discussion}

OpenFerro supports standardized and reproducible lattice-dynamics simulations while preserving flexibility in the choice of Hamiltonian. Researchers can share both a material-specific model and the workflow used to simulate it, without changing the source code. Others can then reproduce the calculation and extend the model using the same numerical tools. This makes individual studies easier to build on and encourages the reuse of validated models.

The nanostructure calculations create an opportunity for quantitative comparisons between lattice and continuum descriptions. Applying both approaches to the same geometry and boundary conditions could establish when local atomistic fluctuations affect the predicted collective patterns and when a continuum description is sufficient. Free-energy sampling offers a further connection: free-energy surfaces obtained from a lattice Hamiltonian can help parameterize continuum models and assess the range over which a chosen analytic form is adequate. These comparisons would make the choice of modeling resolution depend more directly on the process of interest.

Such comparisons require independently validated material models. The simplified elastic and interfacial terms in the present nanostructure examples are sufficient to demonstrate computational capability, but quantitative predictions will require reference data beyond bulk equilibrium states. 

For enhanced sampling, the next challenge is to describe spatially heterogeneous transformations. Macroscopic polarization distinguishes the bulk phases considered here, but different domain configurations can have the same net polarization. Studying nucleation or wall motion will therefore require collective variables sensitive to the relevant spatial structures. 

Further development in terms of scalability will focus on communication in the global Fourier transforms used for dipolar Ewald summation. Efficient multi-node execution would expand the range of accessible sizes and sampling times beyond the single-node benchmarks reported here. 

In parallel, further development of reusable materials libraries is favorable. First-principles calculations and atomistic data, including those generated with machine-learned potentials, can support the construction and validation of lattice models. Sharing these records alongside energy functions would significantly reduce the barrier for first-principles-derived lattice-field simulations.

\section*{Code availability}

OpenFerro is available under the MIT License at \url{https://github.com/salinelake/OpenFerro}. All simulation examples, material parameters, input files, and analysis scripts needed to reproduce the results reported here are publicly released in the same repository. 

\section*{Acknowledgments}
This work benefits from discussions with Yixiao Chen, Ruiqi Gao, Zhi Yao, Andy Nonaka, Xinyu Xu, Yuan Chiang, Lan-Tien Hsu, Karin M. Rabe, Bin Xu, Mingyang Shao, Tomoki Murata, and Ali Ghoreyshi.  During the development of this work, the author used ChatGPT to assist with code optimization and auxiliary scripting. The author retained full technical control over the final codebase and take full responsibility for its accuracy and implementation.
This work was supported by the Laboratory Directed Research and Development (LDRD) Program of Lawrence Berkeley National Laboratory under U.S. Department of Energy Contract No. DE-AC02-05CH11231. This research used resources of the National Energy Research Scientific Computing Center (NERSC), a Department of Energy User Facility, using AI4Sci@NERSC award DDR-ERCAP 34636.

\section*{Author contributions}
P.X. carried out the research and wrote the manuscript.

\section*{Competing interests}
The author declares no competing interests.

\bibliographystyle{naturemag}
\bibliography{ref}

\end{document}